\documentclass[%
reprint,
superscriptaddress,
amsmath,amssymb,
aps,
prb,
floatfix,
a4paper,
longbibliography
]{revtex4-2}

\usepackage{graphicx}
\usepackage{dcolumn}
\usepackage{bm}
\usepackage{hyperref}
\usepackage{gensymb}
\usepackage{xspace}
\usepackage{textcomp}
\usepackage{xcolor}

\begin{document}


\title{Crystal structure evolution induced by the Jahn-Teller effect in mixed-valence silver fluoride Ag$_3$F$_5$}

\author{Dmitry~M.~Korotin}
\email{dmitry@korotin.name}
\affiliation{M.N. Mikheev Institute of Metal Physics of the Ural Branch of the Russian Academy of Sciences, 18 S. Kovalevskaya St., Yekaterinburg, 620108, Russia}

\author{Dmitry~Y.~Novoselov}
\affiliation{M.N. Mikheev Institute of Metal Physics of the Ural Branch of the Russian Academy of Sciences, 18 S. Kovalevskaya St., Yekaterinburg, 620108, Russia}
\affiliation{Department of Theoretical Physics and Applied Mathematics, Ural Federal University, 19 Mira St., Yekaterinburg 620002, Russia}

\author{Yaroslav~M.~Plotnikov}
\affiliation{M.N. Mikheev Institute of Metal Physics of the Ural Branch of the Russian Academy of Sciences, 18 S. Kovalevskaya St., Yekaterinburg, 620108, Russia}

\author{Vladimir~I.~Anisimov}
\affiliation{M.N. Mikheev Institute of Metal Physics of the Ural Branch of the Russian Academy of Sciences, 18 S. Kovalevskaya St., Yekaterinburg, 620108, Russia}
\affiliation{Department of Theoretical Physics and Applied Mathematics, Ural Federal University, 19 Mira St., Yekaterinburg 620002, Russia}

\date{\today}

\begin{abstract}

The silver fluoride Ag$_3$F$_5$ consists structurally of square-planar units formed by four fluoride ions coordinated to a central silver ion, which possesses a partially filled $d$-subshell and the formal valence  of +5/3. 
In this study, we demonstrate that the previously published crystal structure of Ag$_3$F$_5$ is unstable due to the Jahn-Teller effect, arising from the presence of two energetically degenerate $d_{x^2-y^2}$ states sharing a single electron hole. 
Through a full structural relaxation within the DFT+U framework, we identified a new crystal structure for Ag$_3$F$_5$ with reduced symmetry and an energy gain of 151~meV per formula unit relative to the published structure. In this relaxed structure, magnetic chains are formed by silver ions with an electronic hole occupying the $d_{x^2-y^2}$ orbital. 
These results highlight the crucial role of electron correlation effects and related structural distortions in determining the properties of such materials. 

\end{abstract}

\maketitle

\section{Introduction}
Silver fluorides are isoelectronic analogs of high-temperature superconducting copper compounds, a class of materials that has garnered significant attention~\cite{Gawraczynski2019,Grochala2001,Derzsi2022,Liu2022} because of their unique electronic and magnetic properties. Recently, fueled by the growing interest in copper fluorides, the existence of several new silver fluorides, including Ag$_2$F$_5$, Ag$_3$F$_4$, and Ag$_3$F$_5$, has been theoretically predicted~\cite{Rybin2022,Kurzydowski2021} and even synthesized~\cite{Kuder}. These compounds are characterized by the presence of silver ions in mixed-valence states, which is particularly notable from both structural and electronic perspectives.

In this study, we focus on the compound Ag$_3$F$_5$ and investigate its electronic and magnetic structure in detail. This compound is of particular interest not only due to its mixed-valence state, but also because of the potential formation of low-dimensional magnetic structures, such as planes or chains, similar to those found in Cu$_2$F$_5$~\cite{Korotin2021,Korotin2023a}. These structures may lead to significant changes in both the crystal structure and the physical properties of the material. The partially filled electronic $d$-shell of the silver ions highlights the importance of Coulomb correlation effects, which are intertwined with the lattice degrees of freedom~\cite{Leonov2008,Novoselov2016c}.

\begin{figure*}[t]
    \centering
    \includegraphics[width=\linewidth]{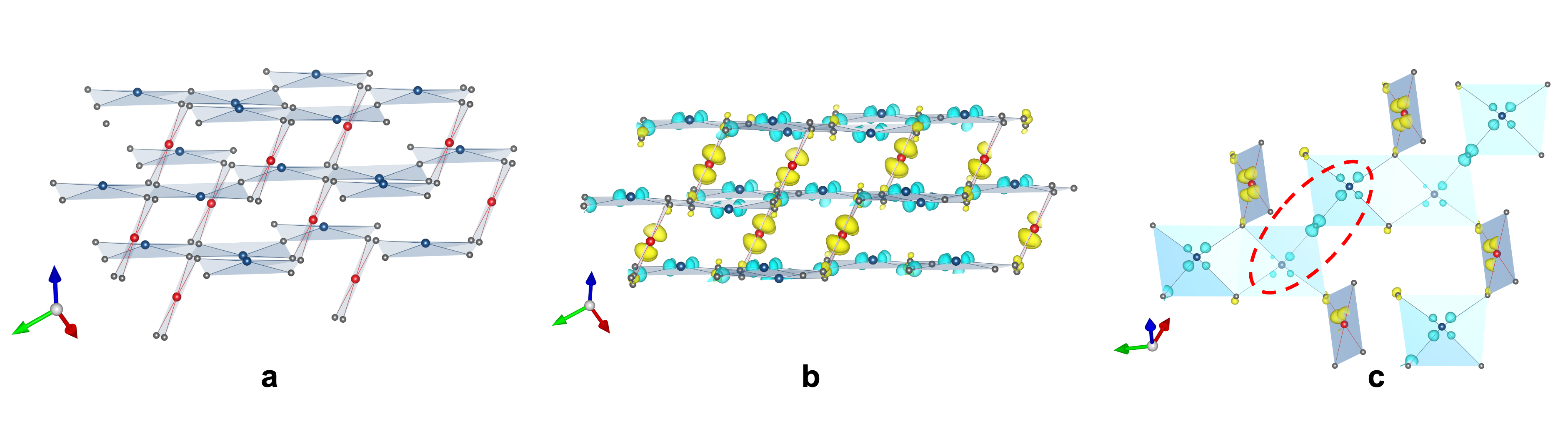}
    \caption{\textbf{(a)} Theoretically predicted crystal structure of Ag$_3$F$_5$ with space group P$\bar{1}$. Red spheres represent Ag ions in the $1b$ Wyckoff position, while blue spheres indicate Ag ions in the $2i$ Wyckoff sites. Small gray spheres denote fluoride ions. \textbf{(b, c)} Side and top views of magnetization density ($\rho^{\uparrow} - \rho^{\downarrow}$) calculated for the P$\bar{1}$ structure of Ag$_3$F$_5$. Yellow regions correspond to positive magnetization, and blue regions to negative magnetization. The red dashed ellipse highlights a region where an electronic hole is shared between two neighboring $d_{x^2-y^2}$ orbitals of Ag ions and the $p$-orbital of the intermediate fluorine ion.}
    \label{fig:start_structure}
\end{figure*}

The formal valence of the silver ion in Ag$_3$F$_5$ is +5/3. As a result, the chemical formula can be written as Ag$^{+}$Ag$^{2+}_2$F$_5$, where the electronic configurations are $5s^04d^{10}$ for the Ag$^+$ ion and $5s^04d^{9}$ for the Ag$^{2+}$ ion, respectively. 
Only a theoretically predicted crystal structure of Ag$_3$F$_5$ is available at the moment~\cite{Rybin2022}. The crystal cell belongs to the symmetry group P$\bar{1}$, with two inequivalent Wyckoff positions for the Ag ions: $1b$~(0,0,$\frac{1}{2}$) and $2i$~(0.3775,0.3163,0.1460). The corresponding structure is depicted in Fig.~\ref{fig:start_structure} (a). Each silver ion is surrounded by four fluoride ions, forming a flat plaquette around the $1b$ position and a slightly convex one near the $2i$ positions. 

Following the multiplicity of the Wyckoff positions, it is reasonable to assume that the nonmagnetic Ag ion, with a $d^{10}$ configuration, occupies the $1b$ Wyckoff position, while the two Ag ions in a $d^9$ configuration are located at the two symmetry-equivalent $2i$ sites.

The square-planar coordination of the ligands induces a splitting of the $d$-orbitals of the ion at the center of the plaquette into the following subshells, listed from highest to lowest energy: $b_{1g} (d_{x^2-y^2})$, $b_{2g} (d_{xy})$, $e_g (d_{zx}, d_{zy})$, and $a_{1g} (d_{3z^2-r^2})$. In Ag$_3$F$_5$, neither of the two types of plaquettes is ideally square: for the Ag1 ion, there are two distinct Ag-F bond lengths (2.14~\AA\ and 2.22~\AA), while for the Ag2 ion, all four Ag-F bond lengths differ (2.15~\AA, 2.16~\AA, 2.24~\AA, and 2.26~\AA). These deviations from an ideal square geometry lift the degeneracy of the $e_g$ subshell of the Ag ions, though the $d_{x^2-y^2}$ orbital remains the highest in energy. It is therefore expected that the electron hole on the Ag2 ion (in the $d^9$ configuration) will occupy this orbital.

These structural and electronic considerations suggest that the compound may exhibit chains of magnetic Ag2 ions in the $d^9$ configuration, with half-filled $d_{x^2-y^2}$ orbitals. Given the importance of electron correlation effects in partially filled $d$-orbitals of transition metals, as well as the potential for different magnetic moment orderings in this compound, we carried out DFT+U calculations to investigate the electronic and crystal structures of Ag$_3$F$_5$. 

Our analysis revealed that the published crystal structure~\cite{Rybin2022} is unstable for symmetry reduction in the lattice. The electronic structure demonstrates features that indicate the presence of an effect analogous to the Jahn-Teller distortion, where the degeneracy of two partially filled $d$-orbitals is lifted through distortions in the crystal lattice, specifically via displacements of the ligands that coordinate the $d$-ions.

\section{\label{sec:methods} Methods}

DFT calculations were performed using the Quantum-ESPRESSO package~\cite{Giannozzi2009}, with pseudopotentials from the Standard Solid-State Pseudopotential Library (SSSP) set~\cite{prandini2018}. The exchange-correlation functional was chosen in the PBEsol form. The energy cut-off for the plane-wave basis set was set to 50~Ry for the wavefunctions and 600~Ry for the charge density expansion. Integration over the Brillouin zone was carried out using a regular $10\times10\times10$ $k$-point mesh.

Electronic correlations were accounted for using the DFT+U method~\cite{Cococcioni2005}, with an effective Hubbard $U$ value of 5~eV~\cite{Rybin2022}, applied to the Ag $d$-orbitals. The convergence criteria for the structural relaxation were set to: total energy less than $10^{-13}$~Ry, total force less than $10^{-5}$~Ry/Bohr, and pressure less than 0.2~kbar. 

Electronic correlations were accounted for using the DFT+U method~\cite{Cococcioni2005}, with an effective Hubbard $U$ value of 5~eV applied to the Ag $d$-orbitals. This value is consistent with previous studies on silver fluorides~\cite{Rybin2022,Derzsi2022,Tokar2021}. The convergence criteria for the structural relaxation were set to: total energy less than $10^{-13}$~Ry, total force less than $10^{-5}$~Ry/Bohr, and pressure less than 0.2~kbar.

Phonon dispersion curves were obtained using finite difference method as implemented in Phonopy~\cite{phonopy-phono3py-JPCM,phonopy-phono3py-JPSJ}. We used $2\times2\times2$ supercell (64 atoms) and $4\times4\times4$ $k$-point mesh for that.

All structures were visualized using the VESTA software package~\cite{VESTA}.

\section{\label{sec:results} Results}

\begin{figure}[t!]
    \centering
    \includegraphics[width=1\linewidth]{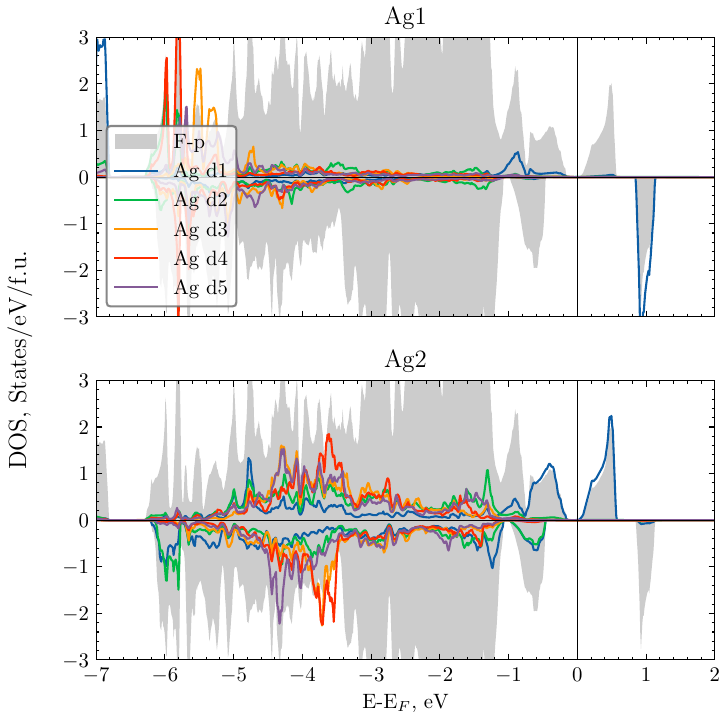}
    \caption{Spin-polarized partial densities of states for the P$\bar{1}$ Ag$_3$F$_5$ crystal structure. The upper and lower halves of each panel correspond to the up- and down-spin channels, respectively. The orbitals $d1$–$d5$ represent atomic states associated with the eigenvectors of the Ag ion occupation matrix, i.e., atomic orbitals defined in the local coordinate system.}
    \label{fig:symm_pdoses}
\end{figure}

The available crystal structure of Ag$_3$F$_5$ was relaxed while preserving the symmetry of the crystal cell, allowing the lattice vectors, angles, and atomic positions to relax until the convergence criteria were met. 
For the relaxed cell, corresponding to the P$\bar{1}$ space group, we compared the total energy of various possible ordering of magnetic moments of Ag ions within the cell and found that antiferromagnetic ordering on the Ag1 and Ag2 types of ions is energetically favorable. 
Calculated partial densities of states (PDOS) for the two types of silver ions, denoted Ag1 and Ag2, are presented in Fig.~\ref{fig:symm_pdoses}. The occupation numbers for the $d$-orbitals of both types of silver ions (Ag1 and Ag2) in the P$\bar{1}$ structure of Ag$_3$F$_5$ are provided in Tab.~\ref{tab:occ1}. 
We define the occupation numbers as the eigenvalues of the site-diagonal $d$-orbital occupation matrices, obtained by projecting the Kohn–Sham states onto atomic-like $d$-orbitals in the local coordinate frame. These eigenvalues ($n^{\uparrow}_i$, $n^{\downarrow}_i$) are listed in ascending order for each spin channel in Tabs.~\ref{tab:occ1} and \ref{tab:occ2}.

The computed densities of states indicate that Ag$_3$F$_5$ behaves as an insulator, with an electronic band gap of approximately 0.22~eV, in agreement with previous findings~\cite{Rybin2022}. Contrary to the initial assumption of a filled $d^{10}$ configuration, Ag1 is found to adopt a $d^9$ electronic configuration. 
The spin-resolved density of states for Ag1 (Fig.~\ref{fig:symm_pdoses}) shows full occupancy in the spin-up channel, while the spin-down $d1$ orbital exhibits a pronounced unoccupied state approximately 1~eV above the Fermi level. The occupation numbers (Tab.~\ref{tab:occ1}) confirm a substantial hole of about 0.56 electrons in this orbital, resulting in a magnetic moment of 0.54~$\mu_\mathrm{B}$. Such spin polarization is inconsistent with a closed-shell $d^{10}$ configuration and instead points to a $d^9$ state, where the hole resides in one of the $d$ orbitals. 

In contrast, the two other silver ions, denoted Ag2, each possess a partially filled $d$-orbital. A hole in the spin-up channel is indicated by a density of states peak at approximately 0.5~eV.

The Ag1 ion has a 0.66 hole in its $d_{x^2-y^2}$ orbital, while the two Ag2 ions collectively share a 0.52 hole. These non-integer values arise from strong hybridization between Ag-$d$ and F-$p$ states, leading to the formation of ligand-hole states: $d^9\underline{L}$ for Ag1 and $\tilde{d}^9\underline{L}$ for Ag2, where $\tilde{d}$ refers to the pair of $d_{x^2-y^2}$ orbitals on neighboring ions.

\begin{table}[b!]
    \centering
    \begin{tabular}{lccccccccccc}
        \hline\hline
        & n$^\uparrow_1$ & n$^\uparrow_2$ & n$^\uparrow_3$ & n$^\uparrow_4$ & n$^\uparrow_5$ & n$^\downarrow_1$ & n$^\downarrow_2$ & n$^\downarrow_3$ & n$^\downarrow_4$ & n$^\downarrow_5$  & $m(\mu_B)$ \\
        \hline
 Ag1 & 0.99 & 0.99 & 1.0 & 1.0 & 1.0 & \textbf{0.44} & 0.99 & 1.0 & 1.0 & 1.0 & 0.54 \\
 Ag2 & \textbf{0.74} & 0.99 & 1.0 & 1.0 & 1.0 & 0.99 & 0.99 & 1.0 & 1.0 & 1.0 & -0.25 \\
    \hline\hline
    \end{tabular}
    \caption{Occupation numbers of the $d$-orbitals in the initial P$\bar{1}$ Ag$_3$F$_5$ crystal structure, for both the spin-up ($n^\uparrow_i$) and spin-down ($n^\downarrow_i$) electronic channels. The magnetic moment of the ion $m(\mu_B) = \sum_i (n^\uparrow_i - n^\downarrow_i)$.}
    \label{tab:occ1}
\end{table}

Since the magnetic moment of the silver ions is primarily associated with a single $d$-orbital, the spatial distribution of electronic holes can be effectively visualized by plotting the isosurface of magnetization density, defined as the difference between spin-up and spin-down charge densities ($\rho^{\uparrow} - \rho^{\downarrow}$) within the unit cell. This approach enables clear identification of hole localization: one component of the spin-polarization corresponds to the hole on the Ag1 ions, while the other component indicates the localization of the hole on the Ag2 ions. The resulting visualization is presented in Fig.~\ref{fig:start_structure}, with panels (b) and (c) showing side and top views of the Ag2 ionic planes, respectively.

The spatial distribution of the magnetization density spin-polarization strengthens our findings regarding electronic hole localization. For the Ag1 ions, the hole is localized on the $d_{x^2-y^2}$ orbital, while for the Ag2 ions, the electronic hole is shared between two neighboring $d_{x^2-y^2}$ orbitals and the $p$-orbital of the intermediate fluorine ion (see the area highlighted by a red dashed ellipse in Fig.~\ref{fig:start_structure}(c)).

At this point, the prerequisite for the structural instability becomes clear. Since the two Ag2 ions occupy the same Wyckoff position in the crystal structure, their energy levels are degenerate by symmetry. This creates a situation with two energetically degenerate $d_{x^2-y^2}$ orbitals hosting one electronic hole. This is a clear indicator of the possibility of a Jahn-Teller effect occurrence that should lift the degeneracy through local distortion of the crystal structure around the silver ion. The instability of the P$\bar{1}$ cell of Ag$_3$F$_5$ within the DFT+U method is also confirmed by the existence of imaginary modes in the calculated phonon spectrum shown in Fig.~S1\cite{supplemental}.

\begin{figure*}[t!]
    \centering
    \includegraphics[width=0.8\linewidth]{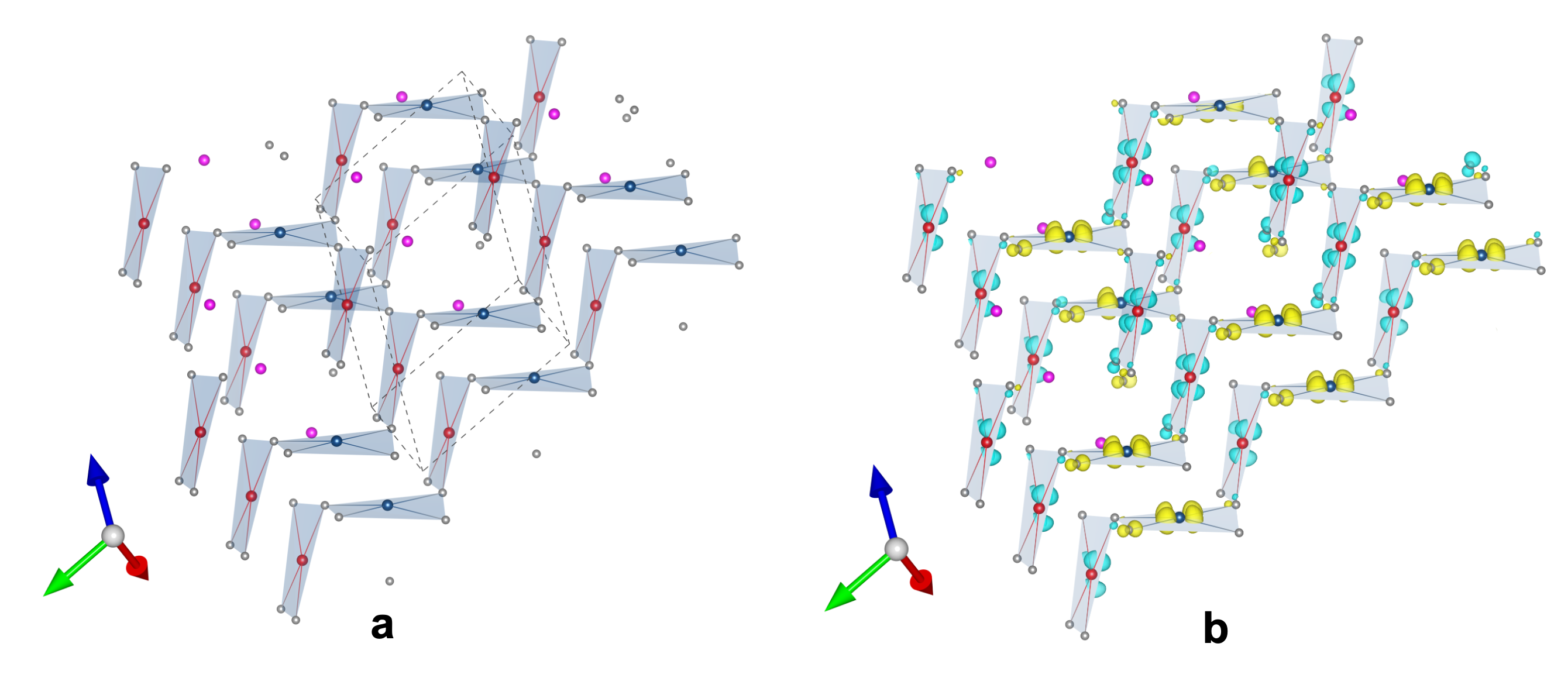}
    \caption{
        \textbf{(a)} Relaxed crystal structure of Ag$_3$F$_5$ with P1 space group. Red spheres represent Ag1 ions, blue spheres indicate Ag2 ions, and magenta spheres shown Ag3 ions (ex. Ag2). Small gray spheres denote fluoride ions. \textbf{(b)} Magnetization density ($\rho^{\uparrow} - \rho^{\downarrow}$) calculated for the P1 structure of Ag$_3$F$_5$. Yellow regions correspond to positive magnetization, and blue regions to negative magnetization.}
    \label{fig:new_struc}
\end{figure*}

\begin{figure}[]
    \centering
    \includegraphics[width=1\linewidth]{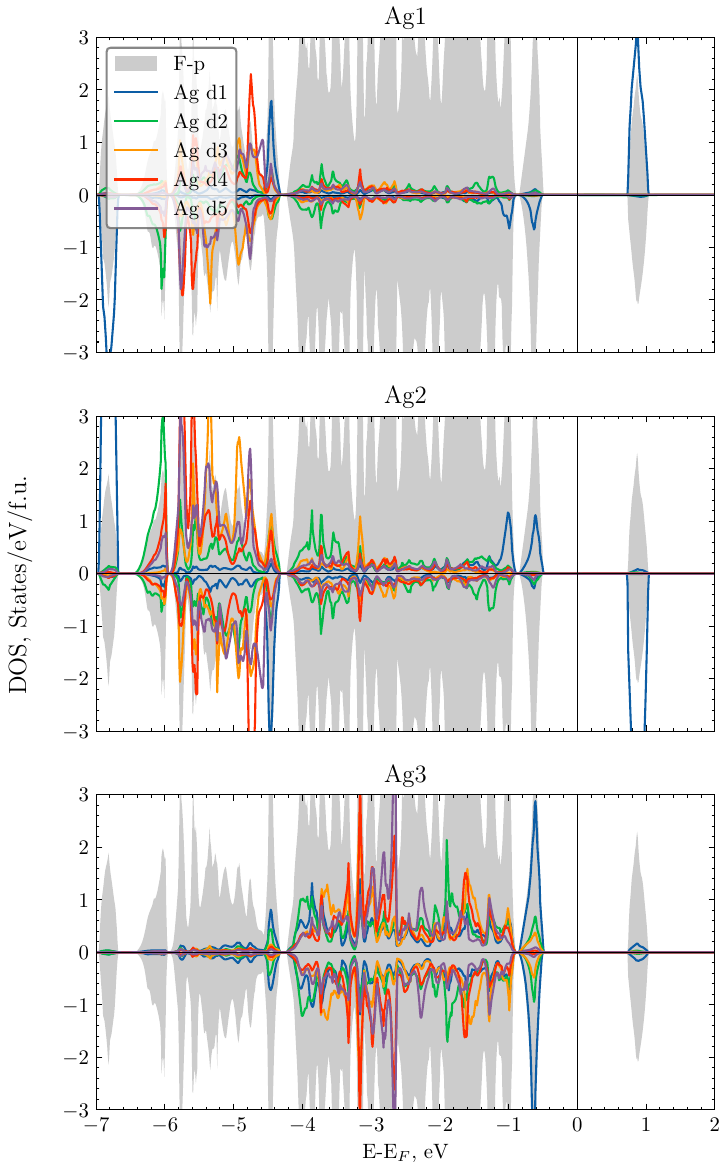}
    \caption{Spin-polarized partial densities of states for the new P1 structure of Ag$_3$F$_5$. The upper and lower halves of each panel correspond to the up- and down-spin channels, respectively. The orbitals $d1$–$d5$ represent atomic states associated with the eigenvectors of the Ag ion occupation matrix}
    \label{fig:nosym_pdoses}
\end{figure}

Guided by the considerations of potential structural instability, we performed a full relaxation of the Ag$_3$F$_5$ crystal lattice, with all symmetry constraints disabled during the process.
This relaxation resulted in a new structure for Ag$_3$F$_5$, characterized by lower symmetry and greater energetic stability. Specifically, the total energy was reduced by 151 meV per formula unit. A visualization of the resulting crystal structure is provided in Fig.~\ref{fig:new_struc} (a) and the structural parameters are in the Supplemental material~\cite{supplemental}. 
The partial density of electronic states and the occupation numbers of the $d$-orbitals for the silver ions in the new structure are depicted in Fig.~\ref{fig:nosym_pdoses} and listed in Tab.~\ref{tab:occ2}, respectively.

\begin{table}[b!]
    \centering
    \begin{tabular}{lccccccccccc}
        \hline\hline
        & n$^\uparrow_1$ & n$^\uparrow_2$ & n$^\uparrow_3$ & n$^\uparrow_4$ & n$^\uparrow_5$ & n$^\downarrow_1$ & n$^\downarrow_2$ & n$^\downarrow_3$ & n$^\downarrow_4$ & n$^\downarrow_5$  & $m(\mu_B)$ \\
        \hline
 Ag1 & \textbf{0.45} & 0.99 & 1.0 & 1.0 & 1.0 & 0.99 & 0.99 & 1.0 & 1.0 & 1.0 & -0.54 \\
 Ag2 & 0.99 & 0.99 & 1.0 & 1.0 & 1.0 & \textbf{0.45} & 0.99 & 1.0 & 1.0 & 1.0 & 0.54 \\
 Ag3 & 0.98 & 0.99 & 1.0 & 1.0 & 1.0 & 0.98 & 0.99 & 1.0 & 1.0 & 1.0 & 0 \\
    \hline\hline
    \end{tabular}
    \caption{Occupation numbers of the $d$-orbitals in the relaxed P1 Ag$_3$F$_5$ cell, for both the spin-up ($n^\uparrow_i$) and spin-down ($n^\downarrow_i$) electronic channels. The magnetic moment of the ion $m(\mu_B) = \sum_i (n^\uparrow_i - n^\downarrow_i)$.}
    \label{tab:occ2}
\end{table}

The relaxed low-symmetry structure of Ag$_3$F$_5$ has the P1 space group: the cell lost the inversion symmetry operation compared to the initial structure. Instead of two types of silver ions, there are now three types: Ag2 ions of the P$\bar{1}$ cell became nonequivalent (noted as Ag2 and Ag3 in the new structure).

The Ag1 and Ag2 ions are each coordinated by four fluorine ions, forming slightly distorted square AgF$_4$ plaquettes. The Ag-F bond lengths within these plaquettes range from 2.05 to 2.17~\AA. These square plaquettes are arranged into zigzag chains along the crystallographic [010] direction, with Ag1 and Ag2 ions alternating within each chain. In contrast, each Ag3 ion is surrounded by six fluorine ions, with Ag-F distances ranging from 2.31 to 2.48~\AA. Thus, in this new structure, the Ag3 ions primarily function as space-filling entities. The covalent bonding between Ag3 and fluorine is expected to be significantly weaker than the bonding between Ag1/Ag2 and fluorine due to the longer bond distances and higher coordination number of the Ag3 ion.

The primary transformation that derives the new P1 structure from the initial P$\bar{1}$ structure is the extension of the fluorine-silver distance for one of the two Ag2 ions. This extension gives rise to the nonmagnetic Ag3-type ion. Consequently, the electronic hole previously shared between two Ag2 ions in the initial cell becomes localized on the new Ag2 ion. This transformation is accompanied by distortion of the remaining Ag-F plaquettes and modification of the translation vectors. The structural evolution is clearly evident when comparing Figs.~\ref{fig:start_structure}(a) and \ref{fig:new_struc}(a).

 The partial densities of states (Fig.~\ref{fig:nosym_pdoses}) and the $d$-orbital occupation numbers (Tab.~\ref{tab:occ2}) indicate that both Ag1 and Ag2 ions exhibit a $d^9$ electronic configuration. These ions display a prominent peak in the density of states at approximately 1~eV above the Fermi level. Additionally, the magnetic moments of the Ag1 and Ag2 ions are antiferromagnetically aligned, as seen from the magnetization density in Fig.~\ref{fig:new_struc} (b), where the formation of antiferromagnetic chains via the $d_{x^2-y^2}$ orbitals is evident. The Ag3 ion, possessing a fully occupied $d$-subshell, is nonmagnetic.

 \begin{figure}[h]
    \centering
    \includegraphics[width=0.9\linewidth]{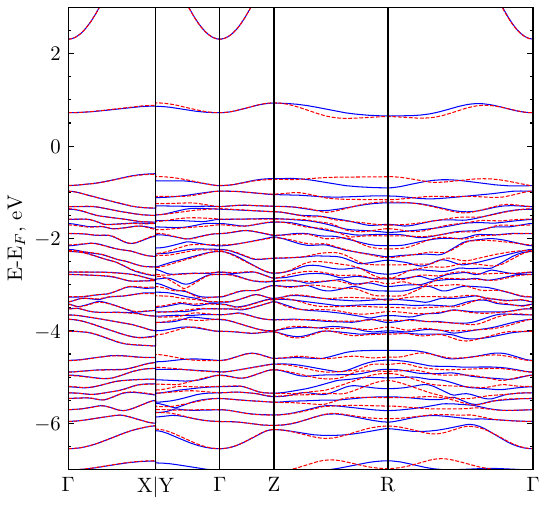}
    \caption{Spin-polarized band structure of Ag$_3$F$_5$, calculated for the new $P1$ structure of Ag$_3$F$_5$.}
    \label{fig:P1_bands}
\end{figure}

The spin-polarized band structure of Ag$_3$F$_5$, calculated for the new $P1$ structure, is shown in Fig.~\ref{fig:P1_bands}. This band structure, with a magnetic configuration having zero net magnetization and finite local moments on two of the three Ag ions (see Tab.~\ref{tab:occ2}), exhibits clear spin splitting along high-symmetry directions $Y \to \Gamma$ and $Z \to R \to \Gamma$. While the absence of spatial symmetries in $P1$ precludes a rigorous symmetry-based classification of the magnetic order as altermagnetic, the observed combination of vanishing total magnetization and spin-resolved electronic structure is characteristic of altermagnetism. These findings suggest that Ag$_3$F$_5$ realizes an \textit{altermagnetic-like} state, highlighting the potential for altermagnetic phenomena even in structurally low-symmetry systems. A more detailed symmetry analysis and exploration of possible spin-orbit-related effects will be necessary to establish the precise nature of this magnetic phase.

\begin{figure}[h]
    \centering
    \includegraphics[width=1\linewidth]{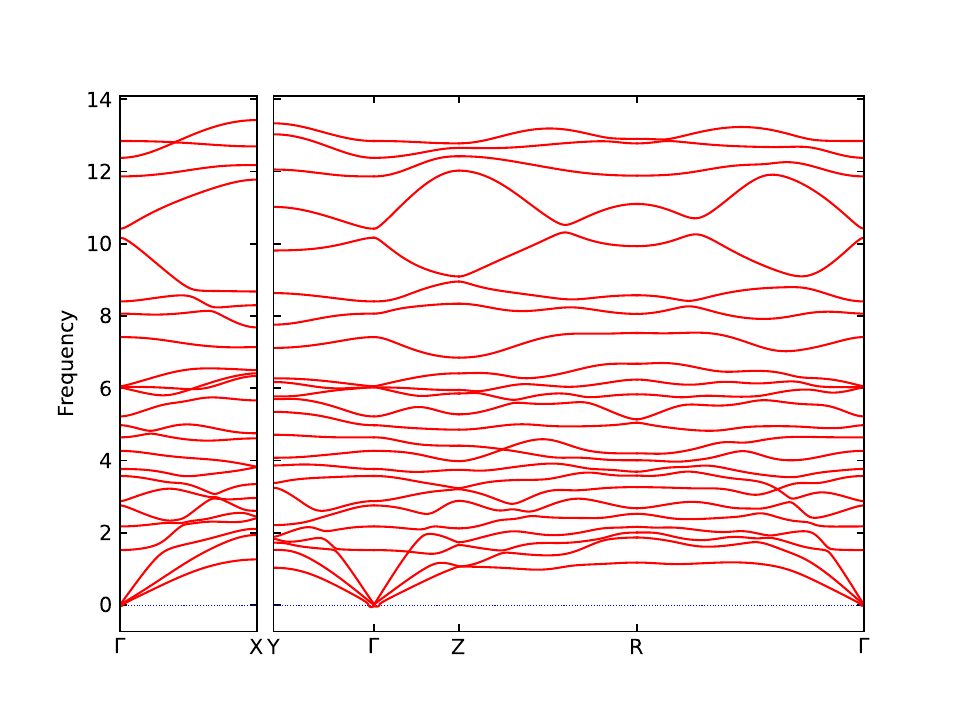}
    \caption{Calculated phonon dispersion curves of the new P1 structure of Ag$_3$F$_5$.}
    \label{fig:nosym_phonons}
\end{figure}

To confirm the stability of the P1 structure of Ag$_3$F$_5$, we calculated the phonon dispersion curves using the finite difference method. The absence of imaginary frequencies (see Fig.~\ref{fig:nosym_phonons}) in the phonon spectrum indicates that the predicted structure is dynamically stable, at least at 0~K.

The energy gap in the new P1 structure of Ag$_3$F$_5$ is significantly larger (1.1~eV) compared to the previously reported P$\bar{1}$ structure (0.2~eV). This substantial increase in the energy gap accounts for the considerable total energy gain observed when the structure relaxes from a higher-symmetry configuration to a lower-symmetry one.

While our calculations focused on Jahn-Teller-driven structural instability, spin-orbit coupling (SOC) may also affect the fine electronic structure of Ag$_3$F$_5$. Silver, as a $4d$ element, has moderate SOC effects that could influence magnetic anisotropy and modify the altermagnetism-like features in the band structure. Though the Jahn-Teller distortions provide a substantial energy gain of 151 meV per formula unit, the interplay of SOC and electronic correlations may be important for fully quantifying the magnetic properties and potential topological features in Ag$_3$F$_5$'s band structure.

\section{Conclusion}
In this work, we conducted a detailed investigation of the electronic and magnetic structure of the recently proposed Ag$_3$F$_5$ compound. The analysis of the electronic density of states and occupation numbers of the $d$-orbitals of the silver ions revealed the presence of two energetically degenerate $d_{x^2-y^2}$ orbitals sharing a single electron hole. Such systems are susceptible to the Jahn-Teller effect, which lifts the degeneracy of the $d$-orbitals through local distortions in the crystal structure. Our findings confirm that Ag$_3$F$_5$ can adopt a crystal structure with lower symmetry, which is energetically more favorable by 151~meV per formula unit compared to the previously reported P$\bar{1}$ structure. In the predicted structure, the two silver ions exhibit a $d^9$ electronic configuration and possess oppositely oriented magnetic moments. Consequently, we predict the formation of chains of antiferromagnetically ordered moments along the [010] crystallographic direction in the new structure of Ag$_3$F$_5$.
The revealed structural stability of Ag$_3$F$_5$ with reduced symmetry and antiferromagnetic order can serve as a starting point for further studies and the search for noble metal fluorides and chlorides with unique electronic and magnetic properties.

\section*{Acknowledgments}
The DFT and DFT+U part of the study were carried out within the state assignment of the Russian Science Foundation (Project 24–12–00024).
The phonon calculations was supported by the Ministry of Science and Higher Education of the Russian Federation (No. 122021000039-4, theme "Electron").

\bibliography{main}

\end{document}